\documentclass[preprint, amsmath]{epl}
\usepackage{epsfig}
\usepackage{graphicx}
\usepackage{bm}

\title{Dichotomy in short superconducting nanowires:
thermal phase slippage vs. Coulomb blockade}
\shorttitle{Dichotomy
in short superconducting nanowires}
\author{A. T. Bollinger($^{\ast}$), A. Rogachev($^{\ast\ast}$), and A. Bezryadin($^{\ast\ast\ast}$)}
\institute{Department of Physics, University of Illinois at
Urbana-Champaign, Urbana, Illinois 61801-3080, USA} \shortauthor{A.
T. Bollinger \textit{et al.}}


\begin{document}

\maketitle

\begin {abstract}
Quasi-one-dimensional superconductors or nanowires exhibit a
transition into a nonsuperconducting regime, as their diameter
shrinks. We present measurements on ultrashort nanowires
($\sim$40-190 nm long) in the vicinity of this quantum transition.
Properties of all wires in the superconducting phase, even those
close to the transition, can be explained in terms of thermally
activated phase slips. The behavior of nanowires in the
nonsuperconducting phase agrees with the theories of the Coulomb
blockade of coherent transport through mesoscopic normal metal
conductors. Thus it is concluded that the quantum transition occurs
between two phases: a ``true superconducting phase" and an
``insulating phase". No intermediate, ``metallic" phase was found.
\end {abstract}

Under certain conditions, usually associated with a critical
resistance per square \cite{Dynes1, Dynes2}, critical total
resistance \cite{Bezryadin1, Bollinger, Zaikin1, Refael1}, or a
characteristic diameter \cite{Giordano1, Lau}, a wire made of a
superconducting metal looses its superconductivity and acquires two
signatures of insulating behavior: i) The resistance increasing with
cooling and ii) a zero-bias resistance peak \cite{Bezryadin1,
Bollinger}. There are many models that capture certain features of
the SIT in 1D wires. Some rely on the ``fermionic" mechanism, in
which disorder combined with electron-electron repulsion suppresses
the critical temperature, $T_c$, to zero \cite{OregFink}. In other,
``bosonic", models the order parameter remains nonzero in the
``insulating" phase while the coherence is destroyed by
proliferating quantum phase slips (QPS) \cite{Giordano2, Zaikin2,
GZQPS, Khlebnikov1, Khlebnikov2, Refael2}. Existing theoretical
models frequently predict a quantum superconductor-insulator
transition (SIT) in thin wires \cite{Zaikin2, Khlebnikov1, Buchler,
Sachdev}, driven, in many cases, by the interaction of the
fluctuating phase with the Caldeira-Leggett environment
\cite{Leggett}. Conditions that make QPS experimentally observable
and the relation of the QPS to the SIT are still being actively
researched \cite{Dynes1, Dynes2, Bezryadin1, Lau, Bollinger,
Zaikin1, Refael1, Giordano1, Arutyunov, Rogachev1, Rogachev2, Chan}.

Here we present a quantitative analysis of the transport properties
of ultrashort nanowires in each of the two phases - the insulating
phase and the superconducting phase. We show that the insulating
phase is characterized by the normal-electron transport and governed
by the Coulomb blockade physics \cite{Nazarov, GZCB1}. The wires in
the superconducting phase exhibit good agreement with the
Langer-Ambegaokar-McCumber-Halperin (LAMH) theory of thermally
activated phase slips (TAPS) \cite{Little, LA, MH}, without any QPS
contribution. The TAPS physics is dominant, even in the vicinity of
the SIT. Thus we conclude that the observed transition occurs
between a truly superconducting phase (which shows no QPS and thus
the resistance approaches zero resistance at $T = 0$) and an
insulating phase in which the wire is in the normal state and the
transport is controlled by weak Coulomb blockade.

\begin{figure}[t]
\begin{center}
\includegraphics[width = 3.3in]{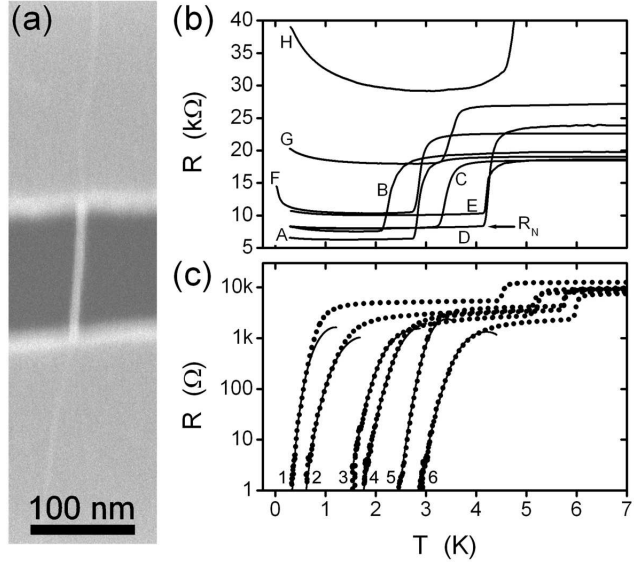}
\caption{\label{fig:RT} (a) SEM micrograph of an $\sim$8 nm wide
nanowire (light) suspended over the trench (black) in SiN. The
sputtered MoGe film was 5.5 nm thick. The white regions at the ends
of the wire indicate that this wire is suspended straight, without
kinking and entering into the trench. The $R(T)$ curves for
insulating and superconducting wires are shown in (b) and (c). The
arrow in (b) shows $R_N$ for sample D. In (c), solid curves indicate
fits to the LAMH-TAPS theory. The fitting parameters are the
coherence lengths, 70.0, 19.0, 11.5, 9.4, 5.6, and 6.7 nm, and the
critical temperatures, 1.72, 2.28, 3.75, 3.86, 3.80, and 4.80 K, for
samples 1-6, respectively. The corresponding normal resistances and
the lengths, determined from the SEM images, are 5.46, 3.62, 2.78,
3.59, 4.29, 2.39 k$\Omega$ and 177, 43, 63, 93, 187, 99 nm
respectively.}
\end{center}
\end{figure}

The nanowires, ranging in length between 43 and 187 nm, are
fabricated by sputtering of amorphous Mo$_{79}$Ge$_{21}$ alloy on
top of suspended fluorinated single-wall carbon nanotubes
\cite{Bezryadin1, Bollinger, Rogachev1}. The wires are homogeneous
as is seen from scanning electron microscope (SEM) images (Fig.\
\nolinebreak \ref{fig:RT}a). The electrodes are deposited during the
same sputtering run as the wire itself. Since Ar-atmosphere
sputter-deposition is isotropic, and due to the small diameter of
the nanotubes ($\sim$1-2 nm), the wires, which occur on the outer
surface of the nanotube, form seamless connections to the
electrodes. The homogeneity of wires is confirmed also by the fact
that their normal resistance is close to that estimated from the
sample geometry and known bulk resistivity ($\sim$200
$\mu\Omega$-cm) \cite{Bezryadin1, Bezryadin2}. Transport
measurements are performed in $^4$He and $^3$He cryostats equipped
with leads filtered against electromagnetic noise \cite{Clarke1}.
One sample (sample F) was measured down to $\sim$20 mK.

Resistance vs. temperature, $R(T)$, data for insulating and
superconducting samples are shown in Figs.\ \nolinebreak
\ref{fig:RT}b and \ref{fig:RT}c, respectively, where $R(T) \equiv
dV/dI$ at $V \rightarrow 0$. The resistive transition observed in
all samples at higher temperatures is that of the film electrodes,
which are connected in series with the wire. As the electrodes go
superconducting, the total sample resistance equals the wire
resistance. The only changing parameter amongst all samples is the
nominal thickness of MoGe (4.0-8.5 nm). Consistently, the critical
temperature of the electrodes decreases gradually as we proceed from
the superconducting sample corresponding to the largest amount of
MoGe sputtered (sample 6) to the insulating sample with the smallest
amount of MoGe sputtered (sample B). The resistance measured
immediately below the film transition is taken as the normal (or
high-temperature) resistance of the wire, $R_N$.

The $R(T)$ curves of superconducting samples are shown in Fig.\
\nolinebreak \ref{fig:RT}c. The fits are made using the LAMH-TAPS
formulas \cite{Rogachev1}. As the resistance starts to sharply drop
with the cooling, the LAMH model becomes valid, and the data exhibit
an excellent agreement with the fits. For the two thinnest samples
(1 and 2), which have the lowest $T_c$'s, the thermodynamic critical
field $H_c(T)$ in the free energy barrier for phase slips had to be
modified: We used the empirical expression \cite{TinkhamBook}
$H_c(T) \propto 1 - (T/T_c)^2$, instead of the usual $H_c(T) \propto
1 - T/T_c$. After such modification the fits matched the data. A
striking result is that our set of ultrashort wires exhibits an
excellent agreement with the TAPS model, without requiring any QPS
contribution \cite{Giordano1, Giordano2}. This is even true for
wires in the vicinity of the SIT. Good agreement with the LAMH model
indicates that the observed superconducting regime is a ``true"
superconducting phase, i.e. the wire resistance is expected to
rapidly approach zero as $T \rightarrow 0$. In this regime, the
wires $R(T)$ exhibits a negative curvature on ln($R$) vs. $T$ plots,
which is an indication that the contribution of QPS \cite{Giordano1,
Giordano2, Arutyunov, Rogachev2, Note} is negligible. Such QPS-free
regime is new and was not seen on longer wires \cite{Giordano1, Lau,
Arutyunov, Chan}.

The $R(T)$ curves of nonsuperconducting wires show a qualitatively
different behavior (Fig.\ \nolinebreak \ref{fig:RT}b). We term them
``insulating" because they reproducibly show (i) an upturn at the
lowest temperatures, i.e. $dR(T)/dT < 0$ (down to $\sim$20 mK, as
was tested for sample F), and (ii) a zero-bias resistance peak, i.e.
$d^ 2V(I)/dI^2 < 0$ for $V \rightarrow 0$ ($I$ is the bias current
and $V$ is the bias voltage).

The observed abrupt change from TAPS to the insulating behavior
strongly suggests that a quantum phase transition does occur in
ultrashort nanowires. This is in contrast to longer wires, which
exhibit a crossover \cite{Giordano1, Lau, Arutyunov, Chan} from a
quasi-superconducting to a quasi-normal regime. We speculate that
this SIT takes place due to coupling of QPS to gapless excitations
in the environment \cite{Zaikin2, Khlebnikov1, Khlebnikov2, Refael2,
Buchler}, similar to the Chakravarty-Schmid transition \cite{Note,
Chakravarty, Schmid, Penttila}. The QPS-free regime can be
understood assuming that the QPS are completely suppressed by a
Caldeira-Leggett environment (e.g. produced collectively by the QPS
cores). In the insulating phase the QPS proliferate and completely
suppress superconductivity, again due to normal cores associated
with each QPS.

\begin{figure}[t]
\begin{center}
\includegraphics[width = 2.4in]{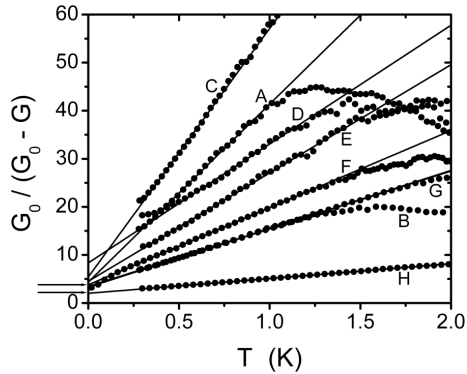}
\caption{\label{fig:GT} Pekola \textit{et al.} \cite{Pekola2}
conductance function $\gamma(T) \equiv G_0[G_0 - G(T)]^{-1}$ plotted
vs. temperature for all insulating samples. The solid lines are fits
to the GZ theory (Eq. (\ref{eqn:GT2})). Values of the fitting
parameter $G_0$ for samples A-H are $G_0^{-1} = $6.14, 7.14, 7.93,
7.76, 9.78, 9.97, 17.33, and 26.10 k$\Omega$, respectively. As
expected, they are close to the corresponding normal resistances of
the wires: $R_N = $ 6.43, 7.54, 8.25, 8.35, 10.33, 10.50, 18.05, and
32.46 k$\Omega$. The predicted range of the offsets is shown by
arrows. The lengths of these wires are 46, 45, 140, 105, 140, 49,
120, and 86 nm.}
\end{center}
\end{figure}

To understand $dR(T)/dT < 0$ seen in our insulating samples, we
consider the theories of the Coulomb blockade (CB) in diffusive
normal wires \cite{Nazarov, GZCB1, Weber1, Weber2}. Nazarov showed
that the CB can survive in a setting in which two plates of a
capacitor $C$ are connected by a homogeneous normal wire (which now
plays the role of a tunnel barrier), even if its resistance is much
lower than the von Klitzing constant, $R_K = h/e^2$, provided that
the wire acts as a coherent scatter \cite{Nazarov}. Golubev and
Zaikin (GZ) \cite{GZCB1} derived useful $I(V)$ formulas, enabling
direct comparisons with experiments. At high temperatures ($k_BT >
E_C$ , where $E_C = e^2/2C$ is the charging energy) the zero-bias
conductance, $G(T)= 1/R(T)$ is
\begin{equation}
\label{eqn:GT1} \frac{G(T)}{G_0} \simeq 1 -
\beta\left[\frac{E_C}{3k_BT} - \left(\frac{3\zeta(3)}{2\pi^4}g +
\frac{1}{15}\right)\left(\frac{E_C}{k_BT}\right)^2\right]
\end{equation}
where $\beta = 1/3$ for diffusive wires.  Also, $\zeta(3) \cong
1.202$, $g \equiv G_0R_K$, and $G_0$ is the conductance in the
absence of the CB. Apart from the second order term $(E_C/k_BT)^2$,
this result is the same as obtained by Kaupinnen and Pekola for a
single tunnel junction \cite{Pekola1}. Originally such result was
derived for a primary thermometer \cite{Pekola2}. The same
expression was derived by Joyez and Esteve \cite{JoyezEsteve} for a
single tunnel junction (i.e. for the dynamic Coulomb blockade),
though in their case the value of parameter $g$ is defined
differently: $G_{JE} \equiv R_K/R_{env}$ ($R_{env}$ is the impedance
of the environment). If the ratio $E_C/k_BT$ is a small parameter,
the diverging terms of the Eq.\ \nolinebreak (\ref{eqn:GT1}) can be
removed by rewriting it as:
\begin{equation}
\label{eqn:GT2} \gamma(T) \equiv \frac{G_0}{G_0-G(T)} \cong
\frac{3k_BT}{\beta E_C} +
\frac{9}{\beta}\left(\frac{3\zeta(3)}{2\pi^4}g + \frac{1}{15}\right)
\end{equation}
In this form the approximate Eq.\ \nolinebreak (\ref{eqn:GT2})
follows closely (as our numerical analysis shows) the exact result
of GZ given in Eq. (28) in Ref. \cite{GZCB1}. In Fig.\ \nolinebreak
\ref{fig:GT} we replot the $R(T)$ data according to Eq.\
\nolinebreak (\ref{eqn:GT2}). The predicted linearity of the Pekola
function $\gamma(T)$ is indeed observed on all insulating samples,
confirming that CB does occur. The parameter $G_0$, adjusted to
produce the best linearity of the curves, matches the high-bias
conductance, as expected (see the caption to Fig.\ \nolinebreak
\ref{fig:GT}). The $E_C$ (Eq.\ \nolinebreak (\ref{eqn:GT2})) is
determined from the slopes. As indicated by Eq.\ \nolinebreak
(\ref{eqn:GT2}), the offset at $T = 0$ should be larger than zero.
The arrows in Fig.\ \nolinebreak \ref{fig:GT} show the expected
range of the offsets, with $g = R_K/R_N$, which is in a fair
agreement with the experiment. At higher temperatures (for example
at $T > 1$ K for sample A) we observe a deviation from the linear
dependence, which might be due to weakening of the proximity effect,
induced by the superconducting leads.

\begin{figure}[t]
\begin{center}
\includegraphics[width = 5.5in]{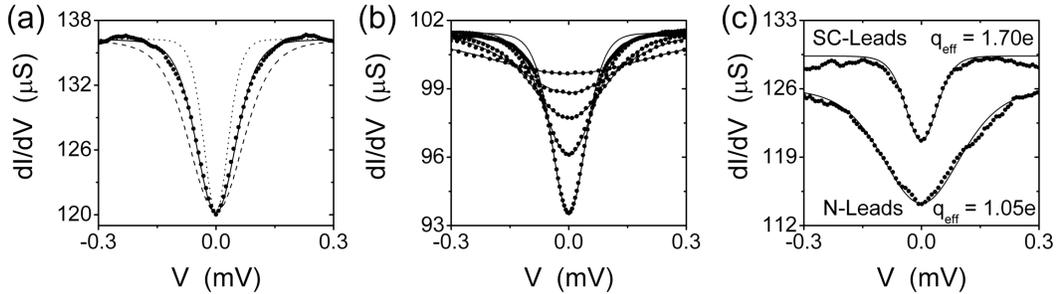}
\caption{\label{fig:GV} (a) The $dI/dV$ vs. $V$ curves for sample B
at $T = 0.28$ K . Comparisons to the GZ theory are shown with
$q_{eff} = e$ (dashed line), $q_{eff} = 1.29e$ (solid line), and
$q_{eff} = 2e$ (dotted line). (b) The $dI/dV$ vs. $V$ curves for
sample E at $T = 0.3$ (deepest dip), 0.5, 0.75, 1.0, and 1.5 K
(shallowest dip). $q_{eff} = 1.53e$. (c) The $dI/dV$ vs. $V$ curves
for sample D at two different magnetic fields. At $B=0$ the leads
are superconducting and $q_{eff} = 1.70e$ whereas at high field ($B
= 9$ T) the leads are driven normal and $q_{eff}$ drops to $1.05e$.
Solid lines are fits to the GZ theory.}
\end{center}
\end{figure}

At higher bias, the CB appears as a zero-bias anomaly (ZBA) observed
in the differential conductance versus bias voltage dependence for
all our insulating wires. The anomaly is similar to the one measured
by Pekola et al. on single-electron transistors \cite{Pekola2}.
Examples of ZBA are shown in Fig.\ \nolinebreak \ref{fig:GV}. We
compare the shape of the $dI(V)/dV$ curves to the GZ theory, in
which the ZBA in the $I(V)$ dependence is given as
\begin{equation}
\label{eqn:IV} I(V) = VG_0 - \frac{e\beta k_BT}{\hbar}
Im\left[w\Psi\left(1 + \frac{w}{2}\right) - iv\Psi\left(1 +
\frac{iv}{2}\right)\right]
\end{equation}
where $w = u + iv$, $u = gE_C/\pi^2k_BT$, $v = eV/\pi k_BT$, and
$\Psi(x)$ is the digamma function. The fitting was done by
differentiating Eq.\ \nolinebreak (\ref{eqn:IV}) and making an
expansion of $\Psi(x)$. The results obtained on high $R_N$ samples
agree well with this prediction. On the other hand, for samples near
the SIT point (i.e. $R_N \approx h/4e^2$ \cite{Note}), the width of
the conductance dip was narrower than predicted (Fig.\ \nolinebreak
\ref{fig:GV}a). Since theoretically the resistance peak becomes
narrower as $v \sim e/T$ increases, the narrowing of the ZBA cannot
be explained by an electronic heating, as this would lead to a wider
anomaly. Understanding this peak-narrowing effect is difficult
because the theory is derived for normal electrodes. We find
empirically then that the $dI(V)/dV$ obtained from Eq.\ \nolinebreak
(\ref{eqn:IV}) can be used to fit our data with $e$ replaced by an
effective charge, $q_{eff}$, which we used as a fitting parameter
along with $G_0$ and $E_C$. With this correction the model matches
the experiments. We speculate that the Andreev reflection taking
place at the ends of the wire leads to $q_{eff} > e$. We also
verified that the $dI(V)/dV$ curves measured at various temperatures
show a good agreement with theory (Fig.\ \nolinebreak
\ref{fig:GV}b). Finally, if a perpendicular magnetic field is used
to drive the electrodes normal (Fig.\ \nolinebreak \ref{fig:GV}c),
the effective charge $q_{eff}$ decreases to $e$, as expected for
all-normal systems. Another explanation for the peak narrowing is
the refrigeration effect \cite{Pekola3}.

\begin{figure}[t]
\begin{center}
\includegraphics[width = 3.6in]{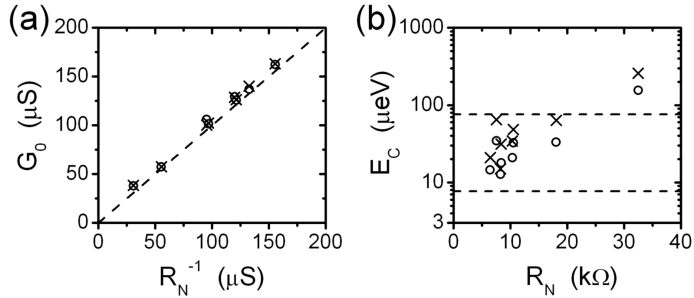}
\caption{\label{fig:Parameters} (a) Values of $G_0$, extracted from
the GZ theory fits, plotted versus $R_N^{-1}$. Symbols are $G_0$
obtained from temperature-dependent ($\times$) and voltage-dependent
($\circ$) data. The dashed line represents $G_0 = R_N^{-1}$. (b)
Charging energies extracted from fits to the GZ theory and plotted
versus $R_N$. Symbols are $E_C$ obtained from the
temperature-dependent ($\times$) and voltage-dependent ($\circ$)
data. Dashed lines indicate the range of $E_C$ estimated for the
geometry of our electrodes.}
\end{center}
\end{figure}

In Figs.\ \nolinebreak \ref{fig:Parameters}a and
\ref{fig:Parameters}b we compare the parameters $G_0$ and $E_C$
obtained from the fits to the $dI(V)/dV$ plots (measured at 0.3 K)
and the same parameters extracted from the fits to the $\gamma(T)$
data. A close agreement is observed, thus confirming the consistency
of our interpretation. Furthermore, it is clear from Fig.\
\nolinebreak \ref{fig:Parameters}a that $G_0$ is very close to
$R_N^{-1}$ for all samples, as it should be.

The $E_C$ extracted from fitting to the $\gamma(T)$ and the
$dI(V)/dV$ plots is shown in Fig.\ \nolinebreak
\ref{fig:Parameters}b as a function of $R_N$. A strong decrease of
$E_C$ with decreasing $R_N$ is seen, possibly due to the
renormalization with $R_N$ \cite{Nazarov}. The exact physical
meaning of the $E_C$ is not well established. One possibility is
that the dynamic Coulomb blockade \cite{JoyezEsteve} that involves
just one barrier (i.e. the entire nanowire) is realized and the
effective capacitance is the capacitance between the electrodes
(i.e. two coplanar thin films \cite{Bezryadin1}). The relevant
sections of the electrodes, which define the effective value of $C$,
are determined by a ``horizon" \cite{Pekola1, Clarke2}. To roughly
estimate the charging energy ($E_C = e^2/2C$) we calculate the
capacitance of the two \textit{coplanar} plates as $C =
\varepsilon\varepsilon_0wK(k)/K(1 - k^2)^{1/2}$, where $\varepsilon
\approx 2$ and $K(k)$ is the complete elliptic integral of modulus
$k = (1 + L/a)^{-1}$ \cite{Song}. For our electrodes of width $w =
10$ $\mu$m, the length $a = 50$ $\mu$m , and the spacing between
closest edges $L = 0.1$ $\mu$m, the capacitance is estimated $C
\approx 1$ fF and the corresponding charging energy is $E_{C,max}
\approx 80$ $\mu$eV (top dashed line in Fig.\ \nolinebreak
\ref{fig:Parameters}b). The lower limit, $E_{C,min}$, was obtained
taking into account the capacitance corresponding to the entire area
of the electrodes, including the contact pads. Using the program
FASTCAP (www.fastfieldsolvers.com) we obtained $E_{C,min} \approx 8$
eV (lower dashed line in Fig.\ \nolinebreak \ref{fig:Parameters}b).
Most of the experimental $E_C$ values fall between the two limits
(Fig.\ \nolinebreak \ref{fig:Parameters}b). However, since the
length of the studied wires is roughly of the same order as the
dephasing length $L_{\phi}$, we note that for wires with $L >
L_{\phi}$ there is another model \cite{GZCB2}, which predicts the
same form of $\gamma(T)$ as Eq.\ \nolinebreak (\ref{eqn:GT2}) but
uses an effective charging energy that depends only on the wire
parameters: $E_C = 3/2\pi^2N_0d^3$ \cite{Golubev}, where $N_0
\approx 8 \times 10^{22}$ (eV cm)$^{-1}$ is the density of states of
MoGe and $d$ is the diameter of a wire. This model gives roughly the
same values of the charging energy.

\begin{acknowledgments}
We thank D.S. Golubev and V. Vakaryuk for important discussions and
R. Dinsmore for helping with the sample fabrication. This work is
supported by NSF CAREER Grant No. DMR 01-34770, DOE grant
DEFG02-91-ER45439 that also supports in part our fabrication
facilities CMM-UIUC.

($^{\ast}$) Present address: Brookhaven National Laboratory, Upton,
NY 11973, USA.

($^{\ast\ast}$) Present address: Department of Pysics, University of
Utah, Salt Lake City, UT 84112, USA.

($^{\ast\ast\ast}$) E-mail: bezryadi@uiuc.edu.
\end{acknowledgments}



\begin{thebibliography}{99}
\bibitem{Note}
The precision with which the SIT critical point equals the quantum
resistance $h/4e^2 \approx 6.45$ k$\Omega$ will be addressed in a
separate publication, after a sufficient number of samples is
tested.
\end{thebibliography}
\end{document}